\documentclass[%
 reprint,
 amsmath,amssymb,
 aps,
]{revtex4-2}

\usepackage{graphicx}
\usepackage{dcolumn}
\usepackage{bm}

\begin{document}

\preprint{APS/123-QED}

\title{Unitarity Problem in Gribov-Zwanziger Theory}

\author{Thitipat Sainapha}
 \email{21wm7101@student.gs.chiba-u.jp}
\affiliation{%
 Department of Physics, Graduate School of Science and Engineering, Chiba University, Chiba 263-8522, Japan
}%


\author{Kei-Ichi Kondo}
 \email{kondok@faculty.chiba-u.jp}
\affiliation{
 Department of Physics, Graduate School of Science, Chiba University, Chiba 263-8522, Japan
}%

\date{\today}

\begin{abstract}
We show that the unitarity problem in the Gribov-Zwanziger theory can be solved by the famous quartet mechanism. The construction of the new quartet in such a theory suggests the introduction of a new form of ghost charge and projection operator. 
\end{abstract}

\maketitle

\section{Introduction}
Unitarity is considered to be one of the important properties of the quantum field theories since it will be not acceptable at all if there is such a case that the considering quantum theory does contain an ill-defined probability. Historically, the quantum Yang-Mills theory constructed through the Faddeev-Popov quantization procedure \cite{Faddeev1967} risks violating the unitarity condition. The particular reason is that the quantization method conditionally introduces anti commuting fields with zero spin, known as ghost and antighost into the theory which violate the so-called spin-statistics theorem. It was shown mathematically that the violation of the spin-statistics theorem will lead to the violation of the unitarity accordingly \cite{streater}. However, the work of Kugo and Ojima showed that the existence of the Becchi-Rouet-Stora-Tyutin (BRST) symmetry in the quantum Yang-Mills theory will smoothly rule out the unphysical degrees of freedom like ghost from the physical Hilbert space. Their method is known as the Kugo-Ojima quartet mechanism \cite{Kugo1978, Kugo1979}.\\
\indent Unfortunately, the discussion of the BRST quartet mechanism excludes the possibility that the Faddeev-Popov method fails to quantize the theory. Theoretically speaking, to quantize the Yang-Mills theory, we have to impose the gauge fixing condition to eliminate the gauge degrees of freedom to make the quantization procedure unambiguous. However, the gauge orbit turns out to cross the gauge fixing hypersurface more than once implying that the gauge redundant degrees of freedom still remain after fixing the gauge. This problem is known famously as the Gribov ambiguity \cite{Gribov1978}. To remedy the problem, we have to force the gauge orbit to be bounded inside the so-called Gribov region. In level of action, it is equivalent to include the extra term into the action of the quantum Yang-Mills theory, the resulting action is called the Gribov-Zwanziger action \cite{Zwanziger1989}. Since the resulting action is non-local, to solve that, we have to add some more sets of auxiliary fields to localize the theory. This attempt causes more serious results since the added fields essentially be the spin-1 anti-commuting fields which again lead to the question about the unitarity of the Gribov-Zwanziger theory.\\
\indent The paper will be organized as follows: In section II, we will explain how the occurrence of the BRST symmetry in the usual quantum Yang-Mills theory will form the BRST quartet which influences the statement that rules out the unphysical degrees of freedom from the theory. In section III, we will take the Gribov ambiguity into account and show that the quartet mechanism works well in the Gribov-Zwanziger model.
\section{BRST quartet in usual quantum Yang-Mills Theory}
Let us start by recalling the occurrence and the impact of the BRST quartet in the usual quantum Yang-Mills theory before we move on to discuss the Gribov-Zwanziger case. The action of the quantum Yang-Mills theory, constructed based on the Faddeev-Popov procedure in the Euclidean space, takes the following form
\begin{equation}
    \begin{split}
        S_{qYM}&=\int d^Dx\;\mathcal{L}_{qYM},\;\;\;\mathcal{L}_{qYM}=\mathcal{L}_{YM}+\mathcal{L}_{gf}\\
        \mathcal{L}_{YM}&= \frac{1}{4}F^a_{\mu\nu}F^a_{\mu\nu}\\
        \mathcal{L}_{gf}&= B\cdot\partial_\mu A_\mu+\bar{c}\cdot\partial_\mu\mathcal{D}_\mu c,
    \end{split}
\end{equation}
where this action is constrained by the Landau gauge condition $\partial_\mu A^a_\mu=0$ to eliminate the gauge redundant degrees of freedom before completing the quantization process. Generically, the theory is no longer invariant under the usual gauge transformation but still invariant under the new symmetry, known as the BRST symmetry, which is the reminiscence of the usual one. According to Noether's theorem, the BRST symmetry transformation is technically generated by the BRST charge $Q_B$ of the form
\begin{equation}
    Q_B=\int d^dx\;\left(\pi_\mu^a(\mathcal{D}_\mu c)^a-\frac{g}{2}f^{abc}\pi^a_c c^b c^c+\pi_{\bar{c}}^a B^a\right),
\end{equation}
where the canonical momenta $\pi_\mu^a$, $\pi_c^a$, and $\pi^a_{\bar{c}}$ conjugate respectively to the Yang-Mills field $A^a_\mu$, ghost $c^a$, and antighost $\bar{c}^a$ are defined as follows
\begin{equation}
   \begin{split}
       \pi_\mu^a&\equiv \frac{\partial \mathcal{L}_{qYM}}{\partial (\partial_0  A^a_\mu)}=F^a_{0\mu},\\
       \pi_c^a&\equiv \frac{\partial \mathcal{L}_{qYM}}{\partial (\partial_0 c^a)}=-\partial_0 \bar{c}^a,\\
       \pi_{\bar{c}}^a&\equiv \frac{\partial \mathcal{L}_{qYM}}{\partial (\partial_0 \bar{c}^a)}=-(\mathcal{D}_0 c)^a.
   \end{split} 
\end{equation}
Let us note that the Hermiticity of ghost $c^a$ and antighost $\bar{c}^a$ are assigned to be $(c^a)^\dagger= c^a$ and $(\bar{c}^a)^\dagger=-\bar{c}^a$ which is the same as the Hermiticity assignment in \cite{Kugo1978,Kugo1979}. The particular reason to do so is to guarantee that $Q_B$ is Hermitian which will be useful later.\\
\indent 
One other thing we have to notice is that the BRST charge carries one ghost number counted by the ghost number operator
\begin{equation}
    Q_c=\int d^dx\;(\bar{c}^a\pi_{\bar{c}}^a-\pi_c^a c^a).
\end{equation}
The fact that the BRST charge does carry exactly one ghost number is expressed mathematically by the following BRST algebra
\begin{equation}
    [iQ_c,Q_B]=Q_B.
\end{equation}
This behavior is theoretically meaningful as we will clarify in what follows. Recall that there are four main unphysical modes in the quantum Yang-Mills theory which are the longitudinal mode of gluon $A_\mu$, Nakanishi-Lautrup field $B$, ghost $c$, and antighost $\bar{c}$. With the simple calculation, we can easily obtain the commutation (graded) relations between these four fields with the BRST operator
\begin{equation}
    \begin{split}
        [iQ_B,A_\mu^a]&=(\mathcal{D}_\mu c)^a,\\
        \{iQ_B,c^a\}&=-\frac{g}{2}f^{abc}c^bc^c,\\
        \{iQ_B,\bar{c}^a\}&=B^a,\\
        [iQ_B,B^a]&=0.
    \end{split}
\end{equation}
The above graded commutation relation truly guarantees that the BRST charge is generically the symmetry generator of the BRST transformation.\\
\indent
From this step, the main objective of the following steps is to ensure that these unphysical modes are not asymptotically observable. To do so, we recall that these fields admit the massless pole structure, so these fields can be written asymptotically as the following expressions
\begin{equation}
    \begin{split}
        A_\mu&\rightarrow \partial_\mu \chi+...\\
        B&\rightarrow \beta+...\\
       \mathcal{D}_\mu c&\rightarrow \partial_\mu \gamma+...\\
        \bar{c}&\rightarrow \bar\gamma+...
    \end{split}
\end{equation}
As a consequence, the graded commutation relation (5) changes accordingly, it casts into
\begin{equation}
    \begin{split}
        [Q_B,\chi^a]&=\gamma^a,\\
        \{Q_B,\gamma^a\}&=0,\\
        \{Q_B,\bar{\gamma}^a\}&=\beta^a,\\
        [Q_B,\beta^a]&=0.
    \end{split}
\end{equation}
Through the canonical quantization, the Fourier modes or the creation operator $(\chi_k^a)^\dagger$, $(\beta_k^a)^\dagger$, $(\gamma^a_k)^\dagger$, and $(\bar\gamma^a_k)^\dagger$ of the asymptotic fields $\chi^a$, $\beta^a$, $\gamma^a$, and $\bar\gamma^a$ satisfy the following commutation relations
\begin{equation}
    \begin{split}
        [Q_B,(\chi^a_k)^\dagger]&=-(\gamma^a_k)^\dagger,\\
        \{Q_B,(\gamma^a_k)^\dagger\}&=0,\\
        \{Q_B,(\bar{\gamma}^a_k)^\dagger\}&=(\beta^a_k)^\dagger,\\
        [Q_B,(\beta^a_k)^\dagger]&=0.
    \end{split}
\end{equation}
These creation operators for asymptotic fields can be legitimately used to construct the states quantum mechanically. Interestingly, these states form the vector space with the BRST complex. As a result, due to the existence of the BRST algebra which is actually the $\mathbb{Z}$-graded Poisson superalgebra, we are mathematically allowed to split the Hilbert space of the considering theory into the direct sum among the vector spaces of each degree characterized by the ghost number. This is equivalent to constructing the projection operator onto the state in the Hilbert space with $n$ ghost numbers \cite{Kugo1979} as the following expression
\begin{equation}
    \begin{split}
        P^{(n)}=&\frac{1}{n}(-\beta^\dagger_k P^{(n-1)}\chi_k-\chi^\dagger_k P^{(n-1)}\beta_k\\
        &+\bar\gamma^\dagger_k P^{(n-1)}\gamma_k+\gamma^\dagger_k P^{(n-1)}\bar\gamma_k).
    \end{split}
\end{equation}
Note that the choice of the projection operator's expression is not totally unique, one might add the term like $-\omega_{kl}\beta^\dagger_k P^{(n-1)}\beta_l$ where $\omega\equiv[\chi_k,\chi^\dagger_l]$ without changing the nature of the projection operator itself.\\
\indent
Significantly, this projection operator can be realized as the exact operator under the BRST transformation. Namely, we can simply write
\begin{equation}
    P^{(n)}=\{Q_B,R^{(n)}\},
\end{equation}
where $R^{(n)}=-\frac{1}{n}(\gamma^\dagger_k P^{(n-1)}\chi_k+\chi^\dagger_k P^{(n-1)}\gamma_k)$. This relation shows that the inner product between the state projected to $P^{(n)}$ and a state with zero ghost number yields zero, implying that the states with non-zero ghost numbers are unphysical by default. This statement really ensures that the physical Hilbert space will not include the particles that violate the spin-statistics theorem; hence, unitarity.

\section{New Definition of Ghost Charge and Quartet in Gribov-Zwanziger Theory}
Let's consider the case in which the usual treatment of the quantization of the Yang-Mills theory fails. If we choose the Landau gauge fixing condition which is the covariant gauge constraint, the Gribov ambiguity will occur consequently. The way to partially solve this problem is to introduce the boundary that cuts off the integral over the gauge space, i.e. the Gribov horizon. The occurrence of the Gribov horizon in the level of action introduces the new action accordingly. Therefore, we have the Gribov-Zwanziger action of the following expression \cite{PhysRevD.92.045039}
\begin{equation}
    \begin{split}
        S_{GZ}=&S_{YM}+\int d^Dx\;(B^h\cdot \partial_\mu A_\mu+\bar{c}\cdot\partial_\mu\mathcal{D}_\mu c)\\
        &+\int d^Dx\;(\bar\varphi_\mu^{ac} M^{ab}\varphi^{bc}_\mu-\bar\omega^{ac}_\mu M^{ab}\omega_\mu^{bc}\\
        &+\gamma^{1/2}_G gf^{abc}A^{h,a}_\mu(\varphi^{bc}_\mu+\bar\varphi^{bc}_\mu),
    \end{split}
\end{equation}
where a set of the auxiliary fields $\{\varphi,\bar\varphi,\omega,\bar\omega\}$ has been added to the theory to make the Gribov-Zwanziger action localized and $M^{ab}$ is a Faddeev-Popov operator defined to be $M^{ab}\equiv -\partial_\mu\mathcal{D}^{ab}_\mu$. These four fields transform as vectors under the Poincare transformation but the latter two fields, i.e. $\omega$, and $\bar\omega$ are spin-one anti-commuting fields, hence, ghost. The existence of these fields would destroy the unitarity of the model since we expect that the same procedure can still be used. First, we have to construct the new nilpotent Noether charge which leaves the localized Gribov-Zwanziger action (7) intact \cite{Kondo:2009qz,PhysRevD.92.045039}. In this particular case, we have
\begin{equation}
    \tilde{Q}=Q_B+Q_\gamma=Q_B+\int d^dx\;(\pi^{ab}_{\varphi\;\mu}\omega^{ab}_\mu+\pi^{ab}_{\bar\omega\;\mu}\bar\varphi^{ab}_\mu),
\end{equation}
where the canonical momenta $\pi_\varphi$, $\pi_{\bar\varphi}$, $\pi_\omega$, and $\pi_{\bar\omega}$ associated to auxiliary fields $\varphi$, $\bar\varphi$, $\omega$, and $\bar\omega$, respectively, are defined below
\begin{equation}
    \begin{split}
        \pi^{ab}_{\varphi\;\mu}&\equiv\frac{\partial\mathcal{L}_{GZ}}{\partial(\partial_0\varphi_\mu^{ab})}=\partial_0\bar{\varphi}^{ab}_\mu,\\
         \pi^{ab}_{\bar{\varphi}\;\mu}&\equiv\frac{\partial\mathcal{L}_{GZ}}{\partial(\partial_0{\bar\varphi}_\mu^{ab})}=\mathcal{D}_0{\varphi}^{ab}_\mu,\\
          \pi^{ab}_{\omega\;\mu}&\equiv\frac{\partial\mathcal{L}_{GZ}}{\partial(\partial_0\omega_\mu^{ab})}=-\partial_0\bar{\omega}^{ab}_\mu,\\
           \pi^{ab}_{\bar{\omega}\;\mu}&\equiv\frac{\partial\mathcal{L}_{GZ}}{\partial(\partial_0\bar{\omega}_\mu^{ab})}=-\mathcal{D}_0\bar{\omega}^{ab}_\mu.
    \end{split}
\end{equation}
We can easily check that the new charge $\tilde{Q}$ of BRST type satisfies the following set of the graded commutation relations
\begin{equation}
    \begin{split}
        [\tilde{Q},\varphi^{ab}_\mu]&=\omega^{ab}_\mu,\\
        \{\tilde{Q},\omega^{ab}_\mu\}&=0,\\
        \{\tilde{Q},\bar{\omega}^{ab}_\mu\}&=\bar\varphi^{ab}_\mu,\\
        [\tilde{Q},\bar\varphi^{ab}_\mu]&=0.
    \end{split}
\end{equation}
However, we have to remark that this charge no longer commutes with the ghost charge in the same as the usual $Q_B$ does. Explicitly speaking, it is simple enough to see that
\begin{equation}
    [iQ_c,\tilde{Q}]=Q_B\not=\tilde{Q}.
\end{equation}
This is a straightforward result we expect to have because the usual BRST charge has nothing related to the new auxiliary fields. To maintain the same structure of the quartet mechanism as those in the Faddeev-Popov quantum Yang-Mills theory, we necessarily introduce the new definition of the ghost number operator as follows
\begin{equation}
    Q_G=Q_c+\int d^dx\;(\bar\omega^{ab}_\mu\pi^{ab}_{\bar\omega\;\mu}-\pi_{\omega\;\mu}^{ab} \omega^{ab}_\mu).
\end{equation}
In this situation, we can easily check that the new ghost number really counts the auxiliary field $\omega$ as a new ghost field and $\bar\omega$ as a new antighost field. Namely, we obtain naturally
\begin{equation}
    \begin{split}
        [iQ_G,\omega^{ab}_\mu(x)]=&-i\int d^dy\;[\pi_{\omega\;\nu}^{cd}(y)\omega^{cd}_\nu(y),\omega^{ab}_\mu(x)]\\
        =&-i\int d^dy\;(\pi_{\omega\;\nu}^{cd}(y)\omega^{cd}_\nu(y)\omega^{ab}_{\mu}(x)\\&-\omega^{ab}_\mu(x)\pi_{\omega\;\nu}^{cd}(y)\omega^{cd}_\nu(y))\\
        =&\;i\int d^dy\;(\pi_{\omega\;\nu}^{cd}(y)\omega^{ab}_\mu(x)\omega^{cd}_\nu(y)\\&+\omega^{ab}_\mu(x)\pi_{\omega\;\nu}^{cd}(y)\omega^{cd}_\nu(y))\\
        =&\;i\int d^dy\;(\{\pi_{\omega\;\nu}^{cd}(y),\omega^{ab}_\mu(x)\}\omega^{cd}_\nu(y))\\
        =&\;i\int d^dy\;(-i\delta^d(x-y)\delta^{ac}\delta^{bd}\delta_{\mu\nu})\omega^{cd}_\nu(y)\\
        =&\;\omega^{ab}_\mu(x),
    \end{split}
\end{equation}
and 
\begin{equation}
    [iQ_G,\bar{\omega}^{ab}_\mu(x)]=-\bar{\omega}^{ab}_\mu(x).
\end{equation}
Similarly, we can obtain the commutation relation between the new ghost number operator and the canonical momenta, we obtain
\begin{equation}
    [iQ_G,\pi_{\omega\;\mu}^{ab}(x)]=-\pi_{\omega\;\mu}^{ab}(x)
\end{equation}
and
\begin{equation}
    [iQ_G,\pi^{ab}_{\bar\omega\;\mu}(x)]=\pi^{ab}_{\bar\omega\;\mu}(x),
\end{equation}
respectively. As a result, we end up with the commutation relation we expect to have
\begin{equation}
    \begin{split}
        [iQ_G,\tilde{Q}]&=Q_B+\int d^dx[iQ_G,\pi^{ab}_{\varphi\;\mu}\omega^{ab}_\mu+\pi^{ab}_{\bar\omega\;\mu}\bar{\varphi}^{ab}_\mu]\\
        &=Q_B+\int d^dx\;\pi^{ab}_{\varphi\;\mu}[iQ_G,\omega^{ab}_\mu]+\int d^dx\;[iQ_G,\pi^{ab}_{\bar\omega\;\mu}]\bar{\varphi}^{ab}_\mu\\
        &=Q_B+\int d^dx\;(\pi^{ab}_{\varphi\;\mu} \omega^{ab}_\mu+\pi^{ab}_{\bar\omega\;\mu} \bar{\varphi}^{ab}_\mu)\\
        &=\tilde{Q}.
    \end{split}
\end{equation}
This commutation relation ensures that the new BRST charge truly carries exactly one ghost number implying that the new BRST charge will behave as the ladder operator among the states of different ghost numbers. Here repeating the same procedure by introducing the asymptotic form of the new fields
\begin{equation}
    \begin{split}
        \omega_\mu&\rightarrow\partial_\mu \Omega+...\\
        \bar\omega_\mu&\rightarrow \partial_\mu \bar\Omega+...\\
        \varphi_\mu&\rightarrow \partial_\mu \phi+...\\
        \bar\varphi_\mu&\rightarrow \partial_\mu \bar\phi+...
    \end{split}
\end{equation}
The superalgebra between the Fourier modes of these asymptotic fields and $\tilde{Q}$ can be written concisely as follows
\begin{equation}
    \begin{split}
         [\Tilde{Q},\phi^{ab}_k]&=\Omega^{ab}_k,\\
         \{\Tilde{Q},\Omega^{ab}_k\}&=0,\\
         \{\Tilde{Q},\bar\Omega^{ab}_k\}&=\bar\phi^{ab}_k,\\
         [\Tilde{Q},\bar\phi^{ab}_k]&=0,
    \end{split}
\end{equation}
and
\begin{equation}
    \begin{split}
         [\Tilde{Q},(\phi^{ab}_k)^\dagger]&=-(\Omega^{ab}_k)^\dagger,\\
         \{\Tilde{Q},(\Omega^{ab}_k)^\dagger\}&=0,\\
         \{\Tilde{Q},(\bar\Omega^{ab}_k)^\dagger\}&=(\bar\phi^{ab}_k)^\dagger,\\
         [\Tilde{Q},(\bar\phi^{ab}_k)^\dagger]&=0.
    \end{split}
\end{equation}
This graded commutation relation implies that we can construct the new set of quartet states legitimately. Once again, the theory mathematically admits the $\tilde{Q}$ Poisson superalgebra, we can consequently construct the projection operator onto the certain state of $n$ ghost numbers as
\begin{equation}
    \begin{split}
        \tilde{P}^{(n)}&=P^{(n)}+P^{(n)}_\gamma\\
        &\equiv P^{(n)}+\frac{1}{n}(\phi^\dagger_k P^{(n-1)}_\gamma\mathcal{O}_1+\bar\phi^\dagger_k P^{(n-1)}_\gamma\mathcal{O}_2\\
        &\;\;+\Omega^\dagger_k P^{(n-1)}_\gamma\mathcal{O}_3+\bar\Omega^\dagger_k P^{(n-1)}_\gamma\mathcal{O}_4),
    \end{split}
\end{equation}
where $P^{(n)}$ is the same projection operator as that defined in the previous section and the $P^{(n)}_\gamma$ is the projection operator contributed by the influence of the Gribov horizon. Since $P^{(n)}_\gamma$ is defined to be the projection operator, it deserves to be one implying that the double action of the projector has to be the same, i.e. $(P^{(n)}_\gamma)^2=P^{(n)}_\gamma$. Thus, we can deduce that $\mathcal{O}_1$ and $\mathcal{O}_2$ are commuting fields while $\mathcal{O}_3$ and $\mathcal{O}_4$ are anti-commuting fields. Nevertheless, more information follows from the fact the Hilbert space equipped with the Poisson superalgebra will be mathematically split into the direct sum of the Hilbert space of each ghost degree, we have
\begin{equation}
    |\psi\big>=P^{(0)}|\psi\big>+\sum_{n=1}^\infty P^{(n)}|\psi\big>.
\end{equation}
If we impose the subsidiary condition $\tilde{Q}|\psi\big>=0$, the projection operator has to commute with $\tilde{Q}$ inevitably. By substituting the relation (26) into the commutation relation
\begin{equation}
    [\tilde{Q},P^{(n)}_\gamma]=0,
\end{equation}
it yields the following equation
\begin{equation}
    \begin{split}
        0=&-\Omega^\dagger_k P^{(n-1)}_\gamma \mathcal{O}_1+\phi^\dagger_k P^{(n-1)}_\gamma \delta\mathcal{O}_1+\bar\phi^\dagger_k P^{(n-1)}_\gamma \delta\mathcal{O}_2\\&-\Omega^\dagger_k P^{(n-1)}_\gamma \delta\mathcal{O}_3
        +\bar\phi^\dagger_k P^{(n-1)}_\gamma \mathcal{O}_4-\bar\Omega^\dagger_k P^{(n-1)}_\gamma \delta\mathcal{O}_4
   \end{split}
\end{equation}
Combining with the fact that the first two operators are commuting fields and the latter two are anti-commuting fields, the possible solution to this equation is as follows
\begin{equation}
    \mathcal{O}_1=-\bar\phi_k,\;\mathcal{O}_2=-\phi_k,\;\mathcal{O}_3=\bar\Omega_k,\;\mathcal{O}_4=\Omega_k,
\end{equation}
where the sign of $\mathcal{O}_1$ and $\mathcal{O}_3$ are not determined directly by the equation (29) but it is chosen to force the projection operator to be Hermitian.\\
\indent 
We can show that the resulting projection operator of degree $n$ is exact under the action of the new BRST charge $\Tilde{Q}$. Namely, we have
\begin{equation}
    P^{(n)}_\gamma=\{\Tilde{Q},R^{(n)}_\gamma\},
\end{equation}
where
\begin{equation}
    R^{(n)}_\gamma=-\frac{1}{n}(\phi^\dagger_k P^{(n-1)}_\gamma\bar\Omega_k+\bar\Omega^\dagger_k P^{(n-1)}_\gamma\phi_k).
\end{equation}
The theoretical meaning of this behavior is as follows: Firstly, if we define the projected state of the ghost number (degree) $n\geq 1$ as $\tilde{P}^{(n)}|\psi\big>\equiv(P^{(n)}+P^{(n)}_\gamma)|\psi\big>\equiv |\psi_0\big>\in \mathcal{V}_0$. The inner product between this state with the state in the total Hilbert space $|\psi\big>\in \mathcal{V}$, as long as the subsidiary condition is implied, turns out to be zero identically
\begin{equation}
    \big<\psi|\psi_0\big>=\big<\psi|P^{(n)}|\psi\big>=\big<\psi|\{\tilde{Q},R^{(n)}\}|\psi\big>=0.
\end{equation}
This implies that the state $|\psi\big>$ and $|\psi+\psi_0\big>$ share the same physical interpretation. Therefore, the physical Hilbert space $\mathcal{H}_{phys}$ is a quotient vector space constructed as the total Hilbert space $\mathcal{V}$ mod the unphysical Hilbert space $\mathcal{V}_0$, i.e.
\begin{equation}
    \mathcal{H}_{phys}\equiv \mathcal{V}/\mathcal{V}_0.
\end{equation}
This fantastic result clearly guarantees that the unitarity property of the local Gribov-Zwanziger theory containing the ghost fields is theoretically safe.
\section{Conclusion and Remark}
We showed that the quartet mechanism works very well in the Gribov-Zwanziger theory by introducing the new definition of ghost charge which follows as a consequence of introducing the new spin-1 anti-commuting auxiliary fields. The interesting question is whether this result hold in the refined Gribov-Zwanziger case or not since, naively speaking, an additional dimension two condensation in the refined Gribov-Zwanziger action might cause a trouble to the new BRST symmetry. This question is important since the refined Gribov-Zwanziger model so far produces several accurate lattice results. Fortunately enough, it has been shown that the generalization of the BRST symmetry can be done properly in the refined case \cite{Sorella:2009vt}. Therefore, there is no problem at all to proceed the same analysis in that model.
\nocite{*}

\bibliography{apssamp}

\end{document}